\documentstyle{l-aa}

\begin{document}


\newcommand{\reff}{\reference}
\newcommand{\equ}{\begin{equation}}
\newcommand{\eequ}{\end{equation}}
\newcommand{\arr}{\begin{array}}
\newcommand{\earr}{\end{array}}
\newcommand{\arry}{\begin{eqnarray}}
\newcommand{\earry}{\end{eqnarray}}
\newcommand{\BF}{\begin{figure}}
\newcommand{\EF}{\end{figure}}
\newcommand{\BI}{\begin{itemize}}
\newcommand{\EI}{\end{itemize}}
\newcommand{\BE}{\begin{enumerate}}
\newcommand{\EE}{\end{enumerate}}
\newcommand{\dis}{\displaystyle}
\newcommand{\BC}{\begin{center}}
\newcommand{\EC}{\end{center}}
\newcommand{\BL}{\begin{flushleft}}
\newcommand{\EL}{\end{flushleft}}
\newcommand{\BTA}{\begin{table}}
\newcommand{\ETA}{\end{table}}
\newcommand{\BT}{\begin{tabbing}}
\newcommand{\ET}{\end{tabbing}}
\newcommand{\TAB}{\begin{tabular}}
\newcommand{\ETAB}{\end{tabular}}
\newcommand{\BD}{\begin{description}}
\newcommand{\ED}{\end{description}}

\newcommand{\RMAA}{{Rev. Mex. Astron. Astrofis. }}

\newcommand{\ApJ}{{ApJ }}
\newcommand{\AsA}{{A\&A }}
\newcommand{\AsJ}{{AJ }}
\newcommand{\Mn}{{MNRAS }}
\newcommand{\Asp}{{Ap\&SS }}
\newcommand{\ApJS}{{ApJS }}
\newcommand{\AsAS}{{A\&AS }}
\newcommand{\JQSRT}{{\em J. Quant. Spectros. Radiat. Transfer}}
\newcommand{\ARAS}{{\em Ann. Rev. Astr. Ap.}}
\newcommand{\Via}{{\em Vistas in Astronomy}} 
\newcommand{\ApJL}{{\em Astrophys. J. Lett.}} 
\newcommand{\Na}{{Nat }}
\newcommand{\AnA}{{\em Ann. d'Ap.}}
\newcommand{\PhR}{{\em Phys. Rev.}}
\newcommand{\Nse}{{\em Nucl. Sci. Engng.}}
\newcommand{\Msait}{{Mem. Soc. Astron. Ital. }}

\newcommand{\Spe}{{Spectral Evolution of Galaxies}}
\newcommand{\RT}{{\em Radiative Transfer}}
\newcommand{\Ntt}{{\em Neutron Transport Theory}}
\newcommand{\Ppim}{{\em Physical Processes in the Intestellar Medium}}
\newcommand{\Iaunu}{{\em IAU Symp.\ 1991}}
\newcommand{\Iauoq}{{\em IAU Symp.\ No.\ 108}}
\newcommand{\Rmcr}{{\em Recent developments of Magellanic Cloud Research}}
\newcommand{\Sn}{{\em Stellar Nucleosynthesis}}
\newcommand{\Iauon}{{\em IAU Symp.\ No.\ 135 1989, }}
\newcommand{\NIM}{{\em Nebulae and Interstellar Matter}}
\newcommand{\Iauno}{{\em IAU Symp.\ No.\  1990, The Galactic and Extragalactic}}
\newcommand{\Lssp}{{\em Light Scattering by Small Particles}}
\newcommand{\AGAS}{{\em Astrophysics of Gaseous Nebulae and Active Galactic
Nuclei}}
\newcommand{\PGAS}{{\em Physics of Thermal Gaseous Nebulae}}
\newcommand{\ND}{{\em Numerical Data and Functional Relationships in Science
and Technology}}
\newcommand{\nrsc}{{\em New Results on Standard Candles}}
\newcommand{\GEO}{{\em Geochim. et Cosmochim. Acta}}
\newcommand{\Pasp}{{PASP }}

\thesaurus{3(11.05.2; 11.16.1; 11.04.1; 11.06.2; 12.03.2)}

\title{$K$ and evolutionary corrections from UV to IR}

\author{B.M.~Poggianti}

\institute{Dipartimento di Astronomia, vicolo dell'Osservatorio 5, 35122
Padova, Italy 
\and Kapteyn Instituut\thanks{\em Present address}, 
P.O. Box 800, 9700 AV Groningen, The Netherlands, bianca@astro.rug.nl}

\date{Received; accepted}

\maketitle

\begin{abstract}

\footnote{Tables 3-41 
are only available in electronic form at the CDS 
via anonymous ftp to cdsarc.u-strasbg.fr (130.79.128.5)
or via http://cdsweb.u-strasbg.fr/Abstract.html.}

$K$ and evolutionary corrections are given for the E, Sa and Sc Hubble types 
for the $U, B, V, R, I, J, H, K$ filters
of the Johnson -- Bessell \& Brett photometric system and 
the $gri$ filters of the modified
Thuan \& Gunn system up to the redshift $z=3$. Their dependence on 
the time scale of star formation in ellipticals is investigated.

The corrections are computed according to an evolutionary synthesis model
that reproduces the integrated galaxy spectrum in the range
1000-25\,000 \AA; such a model makes use of an infrared observed stellar 
library and its results are compared with nearby galaxies.

Evolving spectral energy distributions of
the various Hubble types, as well as optical-IR and IR-IR colour evolution
and adopted filter response functions are also given.


\end{abstract}

\section{Introduction}

The $K$ corrections for galaxies of different morphological types are
necessary to interpret the magnitude-redshift relation, the
luminosity function of galaxies and 
for most of the spectrophotometric studies of distant objects.

The $K$ correction is defined as 
the corrective term that needs to be applied to the observed magnitude
in a certain band due to the effect of redshift.
It does not take into account
the effects of galactic evolution; when this cannot be neglected,
it is necessary to apply a further correction, the evolutionary one (EC),
that can be computed by using spectrophotometric models.
Considering the fact that present observations reach high redshifts
and progressively fainter magnitudes, establishing the connection
between distant and local galaxies requires more and more often the
knowledge of the galactic evolution and the use of both the corrections.

A number of authors have previously published tables of $K$ corrections
(Hubble 1936; Humason et al. 1956; Oke \& Sandage 1968;
Schild \& Oke 1971; Whitford 1971; Oke 1971; Wells 1972; 
Pence 1976; Ellis et al. 1977; Code \& Welch 1979; Coleman et al. 1980;
Frei \& Gunn 1994). In most of the cases these works are limited for one
of more of the following aspects: the number of photometric bands,
the number of galactic types, the maximum redshift considered.
Many of the papers mentioned above only deal with ellipticals, that are
considered the best standard candles at high redshifts.
The biggest efforts to supply an extended set of $K$ corrections have
been made from Pence (1976) and Coleman et al. (1980).
Pence computed the $K$ corrections for the $U, B, V, R$ filters of the 
Johnson system and the $R$ Sandage's filter for the following
morphological types; E/S0, Sab, Sbc, Scd, Im with a maximum redshift of
2.18.
Coleman et al. (1980) found $K$ corrections in the $U, B, V, R$ 
bands for the bulges 
of M31 and M81 and for Sbc, Scd, Im with a $z_{max}=2$.
Frei \& Gunn (1994) have used the energy distribution of Coleman et al.
to compute the $K$ corrections at $z=0.1, 0.2, 0.4, 0.6$
of E, Sbc, Scd and Im for five photometric systems (Johnson $UBV$, 
Gullixson et al. $B_jRI$, Thuan and Gunn $gri$ , $u^{'} g^{'} i^{'} 
r^{'} z^{'} $ and Cousins $RI$).

These studies make use of an empirical method:
with a software programme, the observed spectral energy distribution
of a given morphological type (averaged over a number of objects) is
redshifted. The $K$ corrections are then computed from these 
mean curves and using the filter transmission functions; in this case
there is no need to assume a given cosmological model.
With this method it is obviously not possible to compute the
evolutionary corrections, for which the most direct computing method is
making use of a model of spectrophotometric evolution.

Another advantage of using models instead of observations is that,
in order to cover a wide range of redshift,
the latter often requires the connection  of observations in
various spectral regions, 
most of the times obtained with different instrumentation;
such a connection requires a great accuracy, introduces an
uncertainty and moreover the necessary observations are not always
available.
On the basis of models, Bruzual (1983) calculated the \sl total \rm
corrections ($K$+EC) and the total magnitudes (including $K$, EC and
luminosity distance) for the Johnson's $B, V, K$ bands and the
Koo-Kron $F$ band. Bruzual's model did not include the
most advanced stellar evolutionary phases, that dominate the integrated flux
in the ultraviolet (Post-AGB)
and the infrared (AGB) region. From their spectrophotometric model, 
Guiderdoni \& Rocca- Volmerange (1988)
computed the predicted apparent magnitudes and colours of distant galaxies
for the Johnson $U, B, V, R, I$ filters, Koo-Kron $U^{+}$,$J^{+}, F, N$,
$gr$ from Thuan \& Gunn and some broad-band filters from the Faint Object 
Camera and the Wide Field Camera of the Hubble Space Telescope, taking into
account also the nebular emission and the internal extinction. 
Buzzoni (1995) presented total corrections ($K$+EC) computed from
an evolutionary synthesis model for an elliptical in the Johnson's $B, V, K$
bands and Gunn  $gri$ until a redshift of 1 for different
cosmological parameters.

In this paper both the $K$ and the evolutionary corrections are given
for a set of photometric bands including infrared ones 
($J, H, K$). They are computed from an evolutionary synthesis model and
an attempt has been made to cover a wide range of morphological types
for sufficiently small intervals of redshift up to $z=3$.

\section{Definitions and adopted cosmological parameters}

In this paragraph the $K$ and EC formulae will be presented in detail,
in order to facilitate the use of the following tables. The definitions
that follow are taken from Tinsley (1970).

Consider a galaxy at a redshift $z$, observed at the present epoch $t_0$,
whose light was emitted at the time $t_1$.
Let's define $E(\lambda,t)$ as the monochromatic luminosity
measured at the wavelenght $\lambda$ at the time $t$ in its rest frame,
in units $\rm ergs \, sec^{-1} \AA^{-1} $ or equivalent.

$L_{\lambda_0}$ is defined as the observed luminosity
in the band with effective wavelenght
$\lambda_0$
 (in $\rm ergs \, s^{-1} cm^{-2}$):
\equ
L_{\lambda_0}=\int_0^\infty l_\lambda S(\lambda) d\lambda 
\eequ
where $S(\lambda)$ is the transmission function of the instrument and
 $l_{\lambda}$ is the observed monochromatic flux
($\rm ergs \, s^{-1} \AA^{-1} cm^{-2}$) at the wavelenght $\lambda$. 
Then the following equation is valid:
\begin{eqnarray}
L_{\lambda_0} & = & \frac{1}{4 \pi D^2 (1+z)}\int_0^\infty E(\frac{\lambda}
{1+z},t_1) S(\lambda) d\lambda \nonumber \\  & = &
{\frac{\int_0^\infty E(\lambda,t_0) S(\lambda)d\lambda}{4 \pi D^2 (1+z)}}
\cdot {\frac {\int_0^{\infty} E(\frac{\lambda}{(1+z)},t_0) S(\lambda)d\lambda
}{\int_0^{\infty} E(\lambda,t_0) S(\lambda)d\lambda}} \cdot \nonumber \\
 & & 
\cdot {\frac{\int_0^\infty E(\frac{\lambda}{1+z},t_1) S(\lambda)d\lambda}
{\int_0^\infty E(\frac{\lambda}{1+z},t_0) S(\lambda) d\lambda}} \nonumber \\
\end{eqnarray}
being $D$ the luminosity distance.
Equation (2) can be written as:
\begin{eqnarray}
m_{\lambda_0} & = & M(\lambda_0,t_0) + 5 \log{D} + costant \nonumber \\ 
 & & +
\left[2.5 \, \log (1+z) \, + \, 2.5 \, \log {\frac{\int_{0}^{\infty}
E(\lambda,t_0) S(\lambda) d\lambda}{\int_{0}^{\infty} E(\frac
{\lambda}{1+z},t_0) S(\lambda)d\lambda}} \right] \nonumber \\
 & & + 2.5 \, \log {
\frac{\int_{0}^{\infty}E(\frac{\lambda}{1+z},t_0) S(\lambda) d\lambda}
{\int_{0}^{\infty} E(\frac {\lambda}{1+z},t_1)S(\lambda)d\lambda}} \nonumber 
\\
\end{eqnarray}
The term in the square brackets is the $K$ correction and the last term is
the evolutionary correction.
From Eq.(3) the observed magnitude
$m_{\lambda_0}$ for the band with effective wavelenght $\lambda_0$ 
is equal to the sum of five terms:

a) the absolute magnitude in the same band as it would be measured in the rest 
frame at the epoch of observation $t_0$. This is indicated as
$M(\lambda_0,t_0)$ and corresponds to the numerator of the first term  in 
the Eq.(2);

b) a term that only depends on the luminosity distance $D$;

c) a constant term, depending only on the band used,
that determines the normalization
of the absolute magnitude; 

d) the $K$ correction;

e) the evolutionary EC correction.

The $K$ correction is the difference between the observed magnitude
of the galaxy of age 
$t_0$ measured at the wavelenght ${\lambda}_1={\lambda}_0/(1+z)$
and the magnitude of the same galaxy
of age  $t_0$ computed at 
${\lambda}_0$.
Notice that $t_0$ is the moment of observation
($\sim$ 15 Gyr), while $t_1$ is the time at which the light has been emitted.
Therefore the $K$ correction corresponds to the difference in
magnitude of two objects with identical spectrum due to the redshift:
it does not include in any way the intrinsic evolution of the spectrum
due to the evolution of the stellar populations that contribute to it.

On the contrary, the EC correction depends on the intrinsic evolution
of the spectral energy distribution (SED), 
being the difference between the magnitude of a galaxy of age
 $t_1$ \rm and the same galaxy evolved (whose spectrum is different from
the one of the previous galaxy)
of age $t_0$, both computed at
${\lambda}_1$. It is therefore the difference in absolute magnitude
measured in the rest frame of the galaxy at the wavelenght of emission.

The sum of the $K$ and EC corrections is the difference between
the magnitude of a galaxy of age
$t_1$ redshifted and the one of the evolved galaxy
observed at the time $t_0$ at z=0.
Both corrections are computed on the basis of models of 
spectrophotometric evolution,
assuming a star formation history for each morphological type and having fixed
the cosmological parameters.
In the general case $q_0 \neq 0.5$ it is valid
\begin{eqnarray}
t & = & \frac {-4 q_0}{H_0 {(1-2 q_0)}^{3/2}} \times \nonumber \\ & & \left[
\frac {\sqrt{\frac{1+2 q_0 z}{1-2 q_0}}} {2 \left[ 1- \left(
\frac {1+2 q_0 z}{1-2 q_0} \right) \right]} 
 + \frac {1}{4}
{log}_{e} \left| { \frac {1+ \sqrt{ \frac {1+2 q_0 z} {1-2 q_0} } }
{ 1- \sqrt{\frac {1+2 q_0 z} {1-2 q_0}} } } \right| \right] \nonumber 
\\
\end{eqnarray}
where $t$ is the look-back time,
 $H_0$ is the Hubble constant, $q_0$ is the deceleration
parameter and  $z$ is the redshift.
The tables have been computed 
for  $q_0=0.225$
and $H_0=50 \, \rm Km \, s^{-1} \, {Mpc}^{-1}$, 
corresponding to an age of the Universe $t_0=15$ Gyr.
Corrections for different cosmological parameters can be computed
directly from the evolving SEDs presented in Tables 6-29. 
If one prefers to define the corrections with respect to the \em observed \rm
SED of a given galaxy, the adopted model present-day SEDs given in Tables 3-5
can be replaced with the observed SED.

\section{The spectrophotometric model}

This work makes use of an evolutionary synthesis model that reproduces
the integrated spectrum of a galaxy. A description of this model
is also given in Poggianti \& Barbaro (1996); 
here the essential informations are presented.

The emission of the stellar component is computed with a modified version
of the model by Barbaro \& Olivi (1986, 1989 (BGOF)), that  synthesizes
the SED of a galaxy in the spectral range 1\,000-10\,000 \AA.
The BGOF model includes, besides the main sequence and the central helium 
burning phase, also the advanced stellar evolutionary phases such as 
AGB and Post-AGB. It takes into account the chemical evolution of the galaxy,
therefore the contribution to the integrated spectrum of stellar 
populations of different metallicities.
This model has been successfully employed in the studies of star clusters
in the Magellanic Clouds and of elliptical galaxies (Barbaro \& Olivi 1986, 
1989, 1991; Barbaro 1992; Barbaro et al. 1992).

The stellar evolutionary background has not been changed, while
updates have been made to the library of stellar 
spectra: the new Kurucz stellar atmosphere models (version 1993) have replaced 
the previous ones (Kurucz 1979) and 
the computation of the spectrum has been extended up to
25\,000 \AA. In the infrared region, for stars with  
$T_{eff}>5500$ K Kurucz's models (1992)
have been used. For stars with lower effective temperatures  
the library of observed stellar spectra by
Lan\c{c}on-Rocca Volmerange
(1992, LRV)  has been employed: such spectra cover the spectral range
14\,500-25\,000 \AA, with a resolution between 25 and 70 \AA.
The connection between the optical spectra (1000-10\,000 \AA) and the LRV
spectra has been made by means of black body curves.
For each star the black body temperature has been determined by
imposing that
the resulting colours reproduce the observed ones
(Koornneef 1983 ($V-K$, $J-K$, $H-K$); Bessell \& Brett 1988 ($V-I$, $V-K$, $
J-H$, $H-K$, $J-K$)).

A set of models of the different types has been computed for an age 
15 Gyr and for all the evolutionary times corresponding to various
redshifts according to Eq.(4).
The SFR of an elliptical is approximated with an exponentially decreasing 
function and the average metallicity is assumed solar. Two time scales
have been explored as e-folding times of the SFR: 1 Gyr (E) and 1.4 (E2).
For the spirals, Auddino (1992) galaxy model have been
employed: this is a chemical evolutionary model that includes an inflow
and assumes the SFR to be proportional to the gas fraction.  
This model provides the SFR and the metallicity as a function of time
for galaxies in the type range Sa-Sd. The model parameters for each Hubble type
are determined requiring the model SED to reproduce
the observed colours of local galaxies. The SEDs obtained from this model
reproduce the spectral emission
and absorption characteristics of local galaxies  (Barbaro \& Poggianti 1996), 
as well as their observed average gas fractions.

The inclusion of very advanced stellar phases of extremely metal rich stars
(Greggio \& Renzini 1990) could modify the evolutionary corrections of 
ellipticals for the bluest bands at redshifts  $\geq 0.5$,
because the evolution of the ultraviolet spectral region would be influenced 
by this kind of stellar objects. The $K$ corrections would not be affected,
being the SED of local ellipticals well reproduced by the models also
in the UV range.
Anyway the metallicity of stars in ellipticals is still uncertain: the $\rm 
Mg_2$
index, commonly used to estimate the global metallic content, is difficult
to interpret, due to the excess in early-type galaxies of the ratio
between the  $\alpha$ elements (among which oxygen and magnesium)
 and the iron with respect to the solar value.

Concerning the spirals, the models do not take into account the 
intrinsic extinction due to the presence of dust, that is expected to be 
progressively more significant at increasing redshifts and at decreasing
effective wavelenght. In some cases it will be necessary to consider
a further correction for intrinsic extinction (Di Bartolomeo et al. 1995).

\subsection{Comparison  with nearby galaxies}
 
In principle a first test of the model could be done by comparing the results
with the observations of integrated SEDs of star clusters; good
candidates are the young star clusters in the Magellanic Clouds
(Barbaro \& Olivi 1991). However a great dispersion in the infrared colours of
these objects has been observed; such a dispersion is explained considering 
the stochastic fluctuactions in the mass distribution of the evolved stars
(Barbaro 1992). For this reason and for the uncertainty in the determination 
of the age and the metallicity of each cluster, the comparison has been
made with galaxies, for which the stochastic effects are expected negligible.

\footnotesize
\begin{table*}

\caption[]{Colours of the models of age 15 Gyr}

\begin{flushleft}

\tabcolsep 0.05truecm
\renewcommand{\arraystretch}{0.8}
\TAB{lcccccccccc}
\hline
\noalign{\smallskip}
&$(U-B)$&$(B-V)$&$(V-R)$&$(R-I)$&$(V-J)$&$(V-H)$&$(V-K)$&$(J-H)$&$(J-K)$&$
(H-K)$\\
\noalign{\smallskip}
\hline
\noalign{\smallskip}
El1 & 0.53 & 0.95 & 0.76 & 0.67 & 2.25 & 2.95 & 3.23 & 0.71 & 0.98 & 0.28 \\
El2 & 0.45 & 0.91 & 0.71 & 0.60 & 2.02 & 2.65 & 2.90 & 0.63 & 0.87 & 0.25 \\
El3 & 0.37 & 0.86 & 0.67 & 0.54 & 1.86 & 2.42 & 2.65 & 0.57 & 0.79 & 0.22 \\
Sa &  0.33 & 0.85 & 0.69 & 0.60 & 2.01 & 2.65 & 2.90 & 0.64 & 0.89 & 0.25 \\
Sb &  0.21 & 0.77 & 0.66 & 0.58 & 1.95 & 2.57 & 2.82 & 0.63 & 0.87 & 0.24 \\
Sc &  0.02 & 0.63 & 0.59 & 0.54 & 1.80 & 2.40 & 2.63 & 0.60 & 0.83 & 0.23 \\
Sd & -0.09 & 0.52 & 0.53 & 0.48 & 1.62 & 2.18 & 2.40 & 0.56 & 0.78 & 0.22 \\  
\noalign{\smallskip}
\hline
\ETAB

\end{flushleft}

\end{table*}
\normalsize

Table 1 presents the colours of models of age 15 Gyr of different morphological
types; in the case of the elliptical, the dependence of colours from
the average metallicity is shown: solar (El1, correspondent to the E model), 
Z=0.01 (El2) and Z=0.005 (El3).
Notice that the optical-IR colours ($(V-J)$, $(V-H)$, $(V-K)$) change 
drastically with the Hubble type, while the IR-IR colours ($(J-H)$, $(J-K)$, 
$(H-K)$) change slightly along the type sequence, with differences comparable 
to the observative uncertainty.

\begin{table*}

\caption[]{Comparison models-observations for early-type galaxies; 
*=ellipticals only, without S0}

\begin{flushleft}

\TAB{llclccc}
\hline
\noalign{\smallskip}
                &       & $(U-V)$&$(V-K)$ &$(J-K)$ & $(J-H)$ & $(H-K)$ \\
Bower et al.:   & Virgo & 1.39 & 3.17 & 0.86 &  --     &  --     \\  
                & Coma  & 1.40 & 3.13 & 0.88 &  --     &  --     \\
Persson et al.: & Campo & --   & 3.29 & 0.86 & 0.66  &  0.20 \\                
                & Virgo & --   & 3.22 & 0.86 & 0.65  &  0.21 \\
                & Coma  & --   & 3.19*& --   &  --   &  --   \\
El1             &       & 1.48 & 3.23 & 0.98 & 0.71  &  0.28 \\
El2             &       & 1.36 & 2.90 & 0.87 & 0.63  &  0.25 \\
El3             &       & 1.23 & 2.65 & 0.79 & 0.57  &  0.22 \\
\noalign{\smallskip}
\hline
\ETAB

\end{flushleft}

\end{table*}

Observations in the near-IR have been obtained from  Persson et al. (1979)
for early-type galaxies in Virgo, in Coma and in the field and from
Bower et al. (1992a,b) for a sample of early-type objects in Virgo and in 
Coma.
The average galaxy colours observed by the different authors, corrected for 
redshift , reddening and aperture effects, are compared in Table 2 with
the model results for ellipticals of various metallicities. 
The agreement is satisfactory.
Persson et al. corrected the colours by using Schild \& Oke (1971) and
Whitford (1971) $V$-band $K$-corrections; for the infrared bands, they computed
the corrections using several late-type stars from Woolf et al. (1964).
Bower et al. defined the $U$ and $V$ band $K$ corrections from a series of
template early-type galaxy SEDs, among which those of Coleman et al.,
and the infrared K-corrections from the SED of the K3 giant star $\alpha$ Tau
(Woolf et al. 1964).
For spirals, a great dispersion in the infrared colours
\em within \rm the same morphological type is observed
(Gavazzi \& Trinchieri 1989), therefore spirals are not included in
Table 2; 
the colours of the models for the spirals are inside the range of observed
values.

Moreover the updated evolutionary synthesis model has been used successfully
in the studies of galaxies at intermediate redshifts (Poggianti \& Barbaro 
1995, 1996).

\section{Presentation of the tables and the figures}

From the spectral energy distributions and from the response functions
of the filters in the various bands, the $K$ and EC corrections 
have been computed; they are presented in Tables 31-38.
The corrections have been computed up to $z=3$ for the bands 
$U, B, V, R, I, J, H, K$ of the Johnson's photometric system ($B_1$ and 
$B_2$ corresponding to $B_2$ and $B_3$
from Buser (1978) and $J, H, K$ from Bessel \& Brett 1988) and $gri$ 
of the Thuan \& Gunn system (1976) as modified from Schneider et al. (1983). 
The B magnitude is computed considering the sum of the two filter response
functions $B_1$ and $B_2$ divided by two.
Being 1000 \AA $\,$
the lower limit of the wavelenghts considered from the model,
the U and B bands have been computed respectively up to
$z=2$ and $z=2.5$.

The response function of the filters are given in Tables 39-41; it is useful to
underline that, due to a typing error in Table 4 from Bessell \& Brett,
the definition of the $H$ and $K$ bands can be ambiguous. It is indispensable,
in any case, to check the response function of any filter of interest;
in the case of the $K$ band, the difference between the filter here adopted and
that in \em the figure \rm of Bessell \& Brett can give rise to
errors in the corrections of 0.2 maximum.

The spectral energy distributions of the models of 
different Hubble types of age 15 Gyr
are given in the Tables 3-5 for the E, Sa and Sc
in the spectral range 1012-27\,000 \AA. The SEDs are also presented in Fig. 1.
From top to bottom (at 1000 \AA) the spectra of the Sc, Sa, E2 and E 
are shown; it is visible that the difference
between the spectra of the two ellipticals  is significant only at the shortest
wavelenghts. The rest frame spectra of evolving SEDs of different ages
are presented in Fig. 2 (E), Fig. 3 (Sa) and Fig. 4 (Sc); the ages 
shown are 15, 13.2, 10.6, 8.7, 7.4, 5.9, 4.3, 3.4 and 2.2 Gyr,
corresponding respectively to the redshifts: 0, 0.1, 0.3, 0.5, 0.7, 1.0,
1.5, 2.0, 2.5, 3.0. Such SEDs are given in Tables 6-29, from which the 
interested user can compute any desired property.

Considering the high metal content adopted and the observed 
correlation between the $\rm Mg_2$ index and the absolute luminosity, 
the elliptical model is representative of luminous objects. Due to the
observed substantial variations of the ultraviolet flux with the
galactic luminosity, the results presented here cannot be applied 
to low luminosity ellipticals (i.e. with a lower metal content)
for redshifts  $\geq 0.5$.
Furthermore, the differences between the two model ellipticals
 ($\tau=1$ and  $\tau=1.4$ Gyr) appear significant starting from
 $z=0.6$ ($K$ correction) and $z=0.20$ (EC correction) in the bluest bands.

In order to obtain the observed colour of the progenitor galaxy of
a given type of local galaxy the following relation can be used:

observed colour=colour of the local corresponding galaxy + 
(difference between the $K$ corrections of the first and the second band)+
(difference between the EC corrections of the first and the second band).

This relation can be deduced from Eq.(3).
For instance, if one wants to compute
the expected observed colour (V-J) of an elliptical
at $z=1$:

colour of a local elliptical (V-J)=2.25

$K_V=3.42$, $K_J=0.28$ 

$K_V-K_J=3.14$ 

${\rm EC}_V=-1.87$, ${\rm EC}_J=-0.96$

${\rm EC}_V-{\rm EC}_J=-0.91$ 

expected observed colour$=2.25+3.14-0.91=4.48$.

Figures 5-15 show the $K$ and EC corrections for different bands;
the sudden change in all the curves at $z=2.5$ is due to the fact that the
last two models have been computed with a large redshift step (0.5).
Considering the smooth behaviour of these functions between a redshift
2.5 and 3, a smaller redshift step is not required.

Figures 16-23 show the rest frame and observer's frame colour evolution.

In Table 30 model $K_V$ corrections are compared 
with those of Pence for negligible Galactic extinction.
The differences, starting at relatively low redshift for the
latest types, are partly due to the slightly dissimilar response functions 
adopted and mainly to the differences in the SEDs. 
It must be stressed that two galaxies classified of the same type on the basis 
of their morphological appearance (spiral arms, bulge to disk ratio
etc.) can have significantly different spectra,
indicative of unlike present and past star formation rates.
Therefore the  model galactic sequence should be interpreted as a
``star formation'' sequence, while the results found with an empirical method
will necessarily depend on the single galaxies chosen for the sample.
For this reason the comparison of the two methods results rather difficult.
Furthermore Pence himself defined the ultraviolet observations available to him 
(preliminary OAO data) as ``somewhat uncertain'', especially for E/S0. 
Moreover, due to the lack of ultraviolet 
observations for the Sbc, Pence had to interpolate between the types Sab 
and Scd. Coleman et al.'s results for bulges are in 
agreement with elliptical results from Pence until a redshift $\simeq$0.75 
in the V band. They found instead substantial differences from Pence
in the $K_U$ of the elliptical for $z>0.3$, probably due to the
UV Pence's difficulties mentioned above. A better agreement is obtained 
between their results and the model values presented here.

\section{Summary}

$K$ and evolutionary corrections from UV to IR are presented 
for three kinds of galactic types (E, Sa and Sc) for a number of filters in two
photometric systems. Corrections in other photometric systems are
available by request ( Koo-Kron $U^{+}$,$J^{+}, F, N$  (Koo 1985), 
Cousin $R, I$ (Bessell 1986),
$B_J$, $R_F$ of Couch \& Newell (1980) and 418, 502,  578, 685, 862 of 
the Durham group (Couch et al. 1983)).

These results are based on an evolutionary synthesis model
that successfully reproduces the colours and the shape of the
continuum of the various galactic types of local and distant galaxies.

Spectral energy distributions of the different Hubble types are given
in a wide spectral range (1012-27\,000 \AA) and allow one to compute 
$K$ corrections for any other photometric system.

\begin{acknowledgements}
I wish to express my thanks to G.~Barbaro, for suggestions and corrections
that improved the text, and to A.~Cohen, for helping in typing the tables 
and enjoying life.
\end{acknowledgements}




\begin{figure}
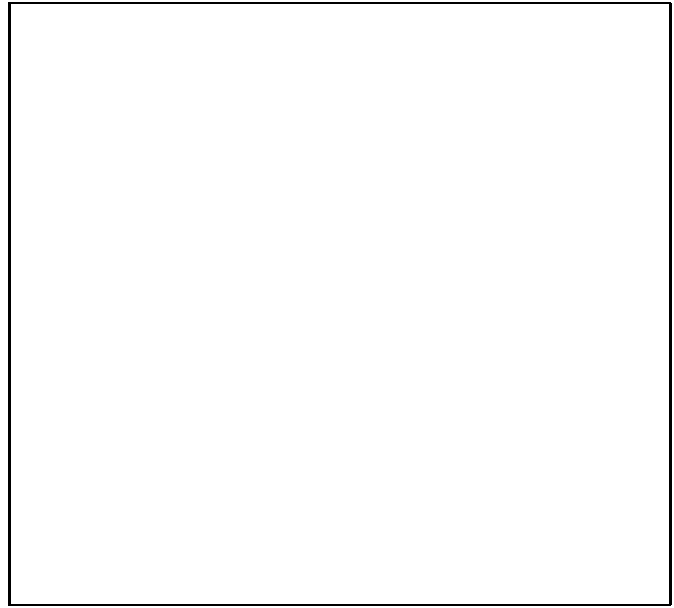

\picplace{8cm}
\caption[]{SEDs of 15 Gyr old models normalized at 5500 \AA. 
From top to bottom (at 1000 \AA):
Sc (dotted line); Sa (short dashed line); E2 (long dashed line); E (solid line)}
\end{figure}

\begin{figure}
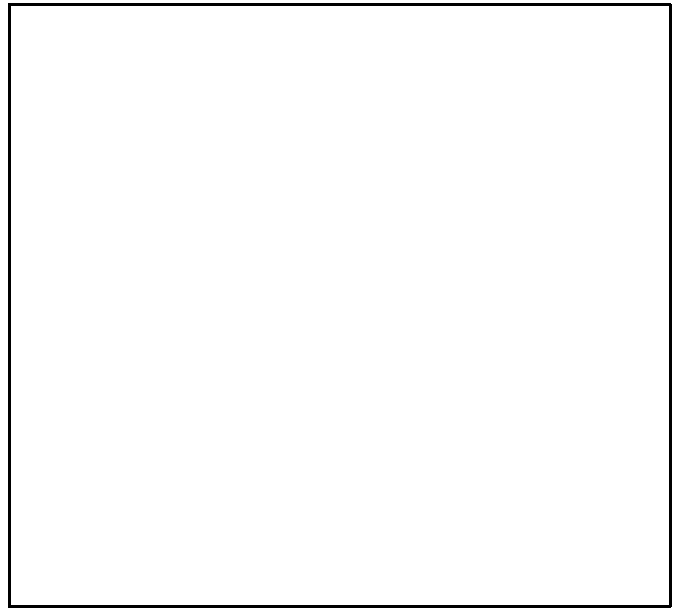

\picplace{8cm}
\caption[]{SEDs of the elliptical model for 9 redshifts: 0, 0.1, 0.3, 
0.5, 0.7, 1.0, 1.5, 2.0, 2.5, 3.0}
\end{figure}

\begin{figure}
\picplace{8cm}
\caption[]{SEDs of the Sa model for the same redshifts of Fig. 2}
\end{figure}

\begin{figure}
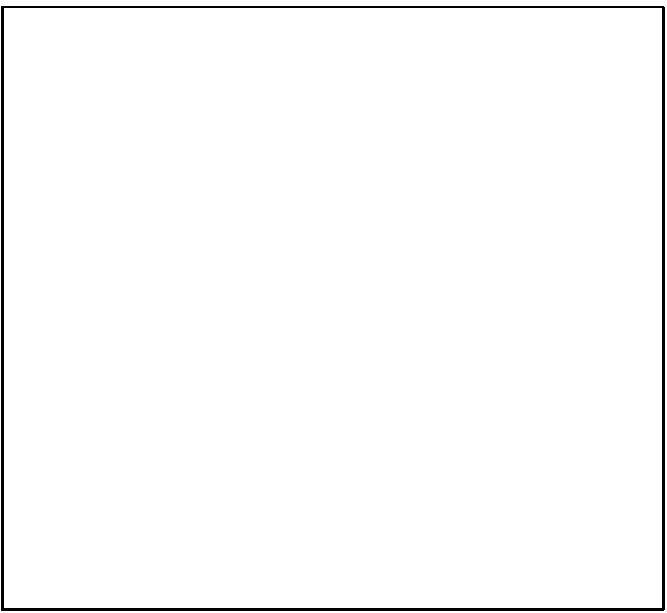

\picplace{8cm}
\caption[]{SEDs of the Sc model for the same redshifts of Fig. 2}
\end{figure}

\clearpage

\medskip
\noindent
{\bf Fig. 5-15. } 
$K$ and EC corrections:
the solid line represents the elliptical with e-folding time 1 Gyr;
the dotted line the elliptical with  $\tau=1.5 $ Gyr (E2); the short dashed 
line refers to the Sa and the long dashed line to the Sc. In some cases the 
curves of the two ellipticals are superimposed and therefore indistinguishable 
\medskip

\noindent
{\bf Fig. 16-23. } 
Rest frame and observer's frame
colour evolution: the former case is
denoted by the ``0'' subscript. Symbols as in Figs. 2-12

\end{document}